\documentstyle[prl,multicol,epsf,aps]{revtex}
\begin{document}
%\draft
\title{
A simple interpretation of quantum mirages  
  }
\author{
  Mariana Weissmann and  H. Bonadeo}
\address{
 Departamento de Fisica,
Comision Nacional de Energia Atomica,
 Av. Gral. Paz 1499 - (1650) San Martin, Prov. de Buenos Aires,
  Argentina.}

\date{\today}
\maketitle

\begin{abstract}

In an interesting new experiment the electronic structure of a magnetic
atom adsorbed on  the surface of Cu(111), observed by STM,
was projected into
a remote location on the same surface.
The purpose of the present paper is to interpret this experiment
with a  model Hamiltonian, using ellipses
of the size of the experimental ones, containing about 2300 atoms.
The charge distribution for the different wavefunctions is analyzed,
in particular, for those with energy close to the Fermi energy of
copper $E_F$.
Some of them 
 show two symmetric maxima
 located
on the principal axis of the ellipse but not necessarily at the foci.
 If a Co atom is adsorbed
at the site where the wavefunction with energy $E_F$ has a maximum
and the interaction is small, the main effect of the adsorbed
atom will be to split this particular wavefunction in two.
The total charge density will remain the same but
 the local density of states will present a dip at $E_F$ at any  site
where the charge density is large enough.
We relate the presence of this dip to the observation of quantum mirages.
Our interpretation
suggests that other sites, apart from the foci of the ellipses,
can be used for projecting atomic images and also indicates the
conditions for other non magnetic adsorbates to produce mirages.
\vskip.1in
Pacs: 73.20.Hb, 73.23.Ps


\end{abstract}

\begin{multicols}{2}

The electronic structure of systems with adsorbed impurities on noble metal
surfaces should depend both on the type and 
the geometric arrangement of
the adsorbed atoms. In particular, the effect of
single transition-metal impurities and also of dimers on the (111)  
surface of  gold  has been studied  recently
with scanning tunneling microscopy \cite{Crommie1,Crommie2,Crommie3}.
Previously, closed loops of impurities had been shown to confine the surface
state wavefunctions of Cu(111), producing what are now called "quantum corrals"
 \cite{corrales}.
In an interesting new experiment the  electronic structure 
surrounding an adsorbed magnetic atom could be projected into a remote 
location
on the same surface \cite{elipses}: a Kondo "signature" in the tunneling
spectrum of a cobalt atom located on one focus of an elliptical corral
was observed as a mirage on the other focus.  
This possibility raises many theoretical questions, the first one
being
the relation between this quantum system and classical image projection.
Other questions regarding the experimental setup are:
How does the appearance of a mirage depend on the 
surface type, on the adsorbed atom type, on the shape of the corral, etc?
The fact that the (111) surface of noble metals presents a symmetry gap
at the $\Gamma$ point and a surface state at about 0.5 eV below the
Fermi energy,  uncoupled from the bulk states, seems of course 
fundamental. For the other questions one must take a closer look and this
is the purpose of the present paper. It must be emphasized, however, 
that we are not attempting a full theoretical description of the 
experiment but rather
a  simplified approach to the previous questions.

We start by assuming a  model Hamiltonian for the interaction of
one adsorbed atom with the surface state of the metal, using the following
approximations:

1. The substrate is represented by a bidimensional array of atoms arranged
in a triangular lattice, representing the (111) surface of copper,
the only effect of the bulk material being to provide the position
of the Fermi level. A tight-binding model with one orbital per site
and first neighbor interactions
is used to represent the atomic interactions within the surface. It
gives an energy band between $E_0-6t$ and $E_0+3t$, that shows a parabolic
shape close to the $\Gamma$ point, as does the surface state in the (111)
surface of noble metals.
The self-energy of the copper atoms ($E_0$) is set to zero and the
nearest neighbor hopping parameter to $t=1eV$,
 a value in the appropriate range for the $s$ band of copper. Since the surface
state at $\Gamma$ is approximately 0.5 eV below the Fermi level in
this case, the resulting Fermi energy is $E_F=-5.5 eV$

2. The corral is simulated by setting a very large
value of $E_0$ for the limiting atoms, or else by just the absence of
atoms beyond the corral limits. We have tried other boundary conditions
and found that there is no qualitative difference in the results.
From previous work \cite{andrea} we know that the surface state develops
when the size of the unperturbed surface region is larger than about 10 lattice
spacings, which limits the size of the smaller ellipses to be considered.

The particular elliptical corral we study here, as 
the experimental one
\cite{elipses}, has semiaxis $a=26d=70 \AA$, $d$ being the nearest neighbor distance
in copper, and
 eccentricity $e=0.5$ so that it
encloses  around 2300 copper atoms.  Diagonalization of the 
tight-binding matrix gives information on the wavefunctions, in particular
 for the relevant levels close
to the Fermi energy. Fig. 1 shows the charge density for four such
levels, from the 39th  to the 42nd eigenfunctions.

 The oscillations observed in the STM topography
\cite{elipses} are expected to be related to the maxima and minima of these 
charge densities.
 It can
be seen that functions number 39 and 42 are both 
quite similar to the pictures in Ref.5  
but one has pronounced maxima very near the foci
of the ellipse while the other has the maxima far away from them.

3. For simplicity,
 the adsorbed atom is assumed to be on the atop position, therefore
interacting with only one of the copper atoms. In the experimental case
the adsorbed atom is cobalt, which is strongly magnetic. When cobalt films
or clusters are deposited on noble metals, a mean field approach to
the electronic structure would postulate that due to a complicated
many-body process the up and down bands separate, one of them being
completely filled up while the other one straddles the Fermi level so as
to be partially occupied \cite{co}. Therefore, the narrow minority band acts as an
impurity with self-energy close to the Fermi energy. In the present case
we only consider the minority level and its   interaction
with the corresponding surface level;
the complicated process is therefore replaced by a resonant
model in which
the fine-tuning of the energies is fundamental. 
This resonant model can also apply to the Kondo resonance at $E_F$.
An atom having a self
energy very close to the Fermi energy is placed atop one of the copper
atoms and the interaction between them is represented by a hopping
constant $t'$.
The value of this parameter is  very important in the model; if it is
too large it leads to a sizable distortion of the surface wavefunctions
which is not the case experimentally. The STM pictures  in Ref.5 show
that one adsorbed cobalt atom inside the ellipse does not change the
topography drastically. If $t'$ is too small its effect is of course negligible.
The value $t'=0.05 eV$ produces what we believe to be an effect related
to the mirage, that is, a splitting of the correct magnitude of
just the level with energy $E_F$,
to which the
atop atom is tuned. The wavefunctions in a tight-binding model are
obtained as a linear combination 
of the atomic orbitals and elementary considerations show
that the splitting is
directly proportional to $t'$ and to the magnitude of the particular 
coefficient corresponding to the copper atom that is connected to the
adsorbed atom, and inversely proportional to the mistuning of the energy.

Some results obtained from this  model Hamiltonian are shown in Fig. 2.
Fig. 2a shows a  projection of the probability density  from state 42 of
Fig. 1, while Figs. 2b and 2c are the probabilities of the two split levels 
for the case
of an adsorbed atom  placed atop the site $x=12d$
where the wavefunction is maximal, near a focus of the ellipse. 
A similar plot is obtained when the adsorbed atom is placed in a  position
where there is a smaller secondary maximum, as for example $x=7d$. The only difference between
the two situations
is the larger energy splitting in the first case (8meV versus 5meV).
 The difference between Figs. 2b and 2c is 
 a consequence of the slight mistuning of the adatom self-energy and the Fermi energy.
The surface topography will remain unchanged by the interaction, as the sum
of the split wavefunctions is almost identical to the original one, but
a spectrum scanning the density of states will find a dip at the Fermi level
of the order of the energy splitting, less than 0.01 eV. This dip will appear
in the local densities of states proportionally to the
corresponding coefficients and therefore will be larger at the foci. 
A similar calculation on function 41, with a close by energy, placing the atop atom on the
position of the wavefunction maximum (that is, far away from the focus of
the ellipse) gives a similar result that should be 
observed at the symmetrically opposed site.
Also, a small change in the size of the ellipse shifts the eigenvalues so that the
Fermi energy may tune to wavefunction number 39, that has two maxima on the principal
axis of the ellipse but far away from the foci and therefore for this ellipse
the mirage would not appear at the foci.

In fact, in this model there is nothing special about the foci of the
ellipse.  If you pick the right size and
eccentricity for it so that the surface eigenfunction with energy $E_F$
has only two large symmetric maxima the splitting will also be largest. If the ellipse
does not have a wavefunction at $E_f$ with such characteristics, no effect will
be observed within our model. In particular, if we modify the size of the ellipse
keeping the eccentricity constant, we observe alternate appearance and disappearance
of wavefunctions of the described type with maxima near the foci at appropriate
energies,
 in agreement with experiment\cite{elipses}.

This leads to the question: why ellipses? Any other corral with
wavefunctions having a small number of well defined, pronounced,
maxima should be as good for producing mirages.
However, although we have explored several other possible  shapes (ellipses
narrowed at their center, for instance),
ellipses seem to be the most efficient in this sense. 

A final question : Is magnetism  essential for obtaining a mirage? Cobalt
is obviously an ideal atop atom because its local density
of states develops a very narrow feature just at $E_F$: it 
 fine-tunes itself; this is not
the case for Fe, Ni, Cr, etc. However,
other atoms and even non-magnetic ones may also have large densities of
states near the Fermi level of the substrate, for example, S on Pd (111)
\cite{s}. As it was reported that no mirage effect appeared for S or CO
on Cu(111) \cite{elipses} we decided to perform an ab-initio
calculation to try to understand why this is so.

In a  previous paper \cite{auco} we calculated  the electronic 
densities of states of a periodic system that simulates the  experimental
situation of an adsorbed atom on a noble metal surface. It 
consists of repeated slabs of five layers of Au(111) with 
Co atoms deposited on both sides of them 
forming a dilute adsorbed layer. For this purpose we
used the FP-LAPW method and the LSDA approximation 
with the WIEN97 code \cite{Blaha}. The main result of that paper was that the
minority spin contribution to the density of states due to the $3d$ orbitals
of Co with  $m=0$ 
  is a very narrow band
 precisely at Ef.

In the present work we have repeated the same type of calculation
for other cases, such as S and Si on Cu(111). The
unit cell  is the same as in the previous paper \cite{auco};
it has five  noble metal layers with three atoms in each layer. One
impurity atom is located on each side of the slab at one of the hollow
sites of the (111) slab structure forming a $\sqrt{3}\times\sqrt{3}$ 
adsorbed layer.
The slabs are  separated by enough empty space so as to
simulate non interacting surfaces.  The distance between impurity
atoms on the same plane is 5 \AA. As in this calculation we are
intending to simulate the real material, we used the more probable
position rather then the atop one.  

For S on Cu(111) we found that the local  density of states
is not peaked at $E_F$ but at slightly  lower energies, so that if
our interpretation is correct a
bias voltage of 0.3 eV would help observing the mirage effect. 
If instead of S one  would deposit
Si, the tuning to the Fermi energy should be  better (see Fig. 3). 
Because of the symmetry of the surface state, the orbitals of the 
adsorbed atom that should interact with it are those with $l=1,m=0$.
These show a narrow feature close to $E_F$, although not as narrow
as that of Co.
 It must be noted that  this calculation does not intend to study
the interaction of the adsorbed atom with the surface state, which
should not be present due to the periodic location of the
adsorbed impurities, but only the interaction with the bulk
noble metal that determines the position of the adsorbed atom energy
with respect to the Fermi energy.

{\bf Conclusions}

 We have performed a  calculation which 
gives some insight on the main features of the imaging effect of the
elliptical corral and  leads to  suggestions for future experiments
which could confirm - or not- the  idea presented in this paper.
It should be possible to make use of sites other than the foci of 
the ellipses to project atomic images. In particular, while an ellipse
of $e=0.5$ and $a=26d$  has the interesting wavefunction with its
maxima near the foci ($x=12d$  ), reducing the size to $a=25d$  and keeping $e=0.5$
 produces
a wavefunction of a  similar shape but with its maxima far from the foci, at $x=16d$
at $E_F$.
In this sense the quantum mirage appears to be quite different from
 classical  focusing by an ellipse.
Also,  other atoms such as S and Si
could be used, the fine tuning being achieved by proper adjustment of
the voltage bias in the STM experiments.

\newpage
\begin {figure}
\caption{ Charge densities for the quantum corral wavefunctions number
39 to 42. The brightness indicates increasing charge density.
The eigenvalues in eV are e(39)=-5.5581, e(40)=-5.5578, e(41)=-5.5512, 
e(42)=-5.5219.
Wavefunctions 39 and 42 show two strong symmetrical maxima on the principal 
axis
at $x=16d$ and $x=12d$ but also smaller symmetrical maxima in other places. 
}
\label{Fig. 1}
\end{figure}

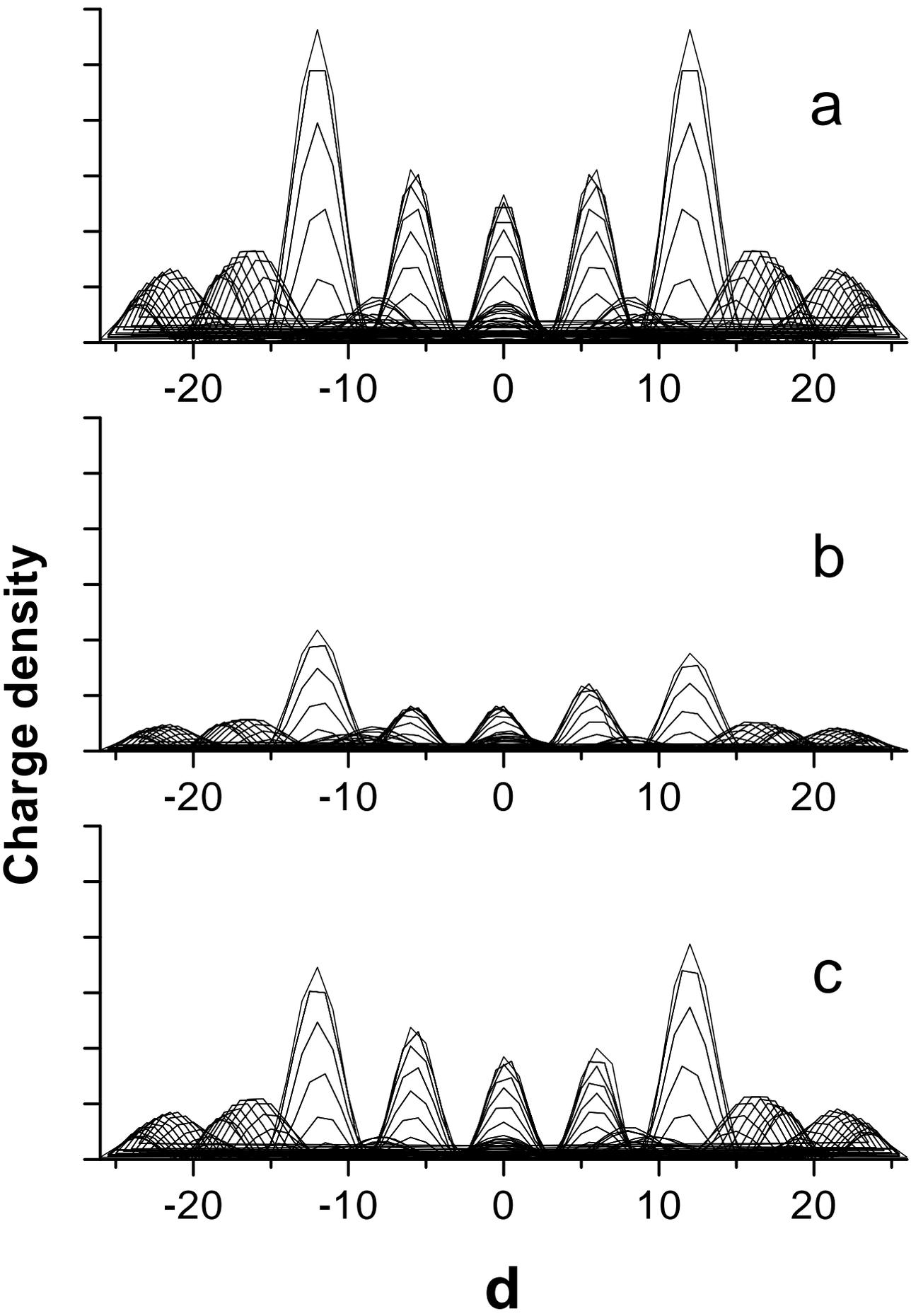
\begin {figure}
\caption{ (a) Projection of the charge density for the quantum corral wavefunction number
42  and (b),(c) corresponding split charge densities when an adsorbed atom is added
at $x=12d$. The x axis is the principal axis of the ellipse while the y
axis contains the charge density of all atoms with the same coordinate x.}
\label{Fig. 2}
\end{figure}

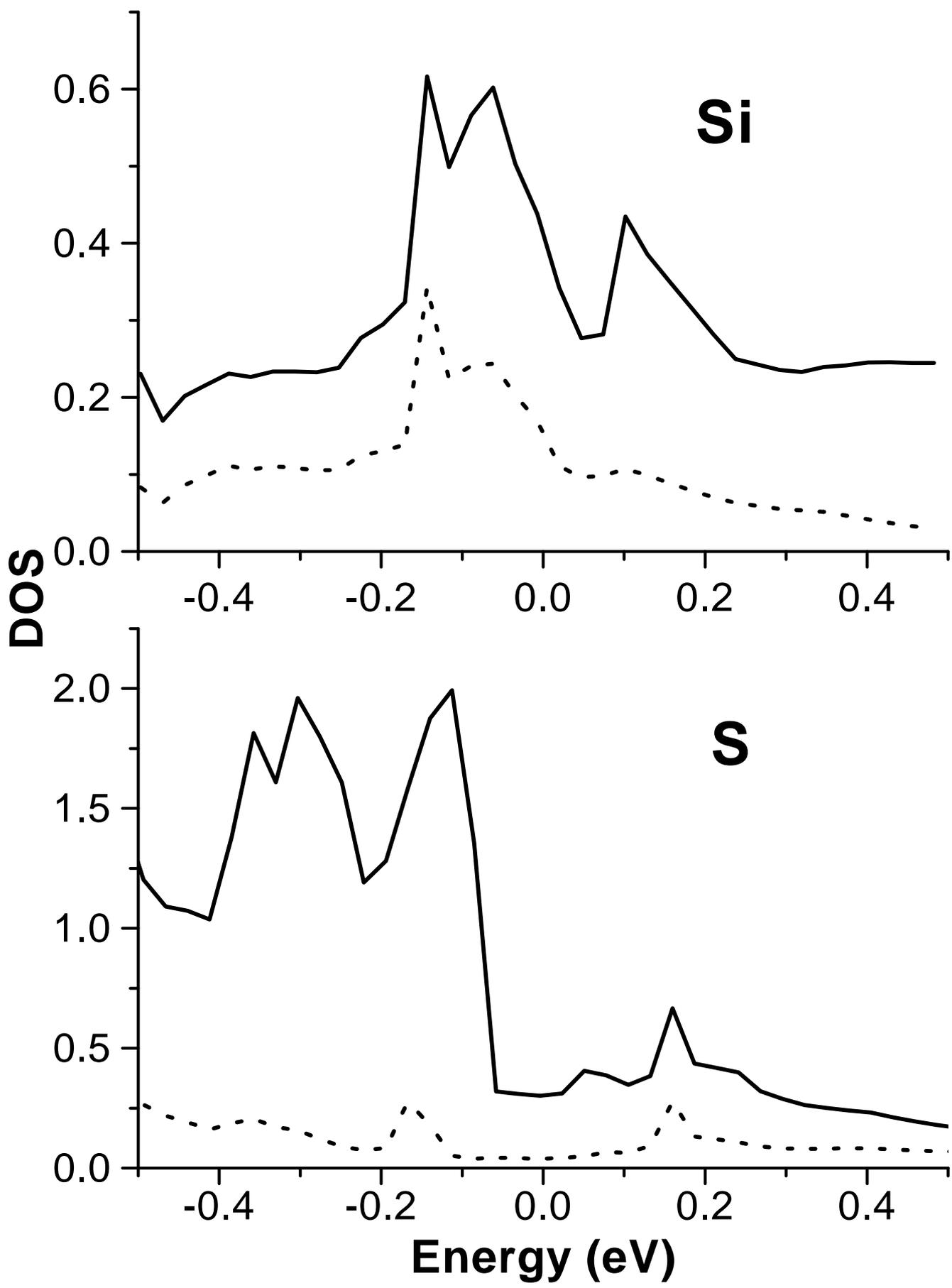
\begin {figure}
\caption{ Full line: Local density of states at the adatoms S and Si adsorbed
on Cu(111), close to the Fermi energy; Dotted line: Only $m=0$ states (orbital $p_z$).} 
\label{Fig. 3}
\end{figure}

\end{multicols}
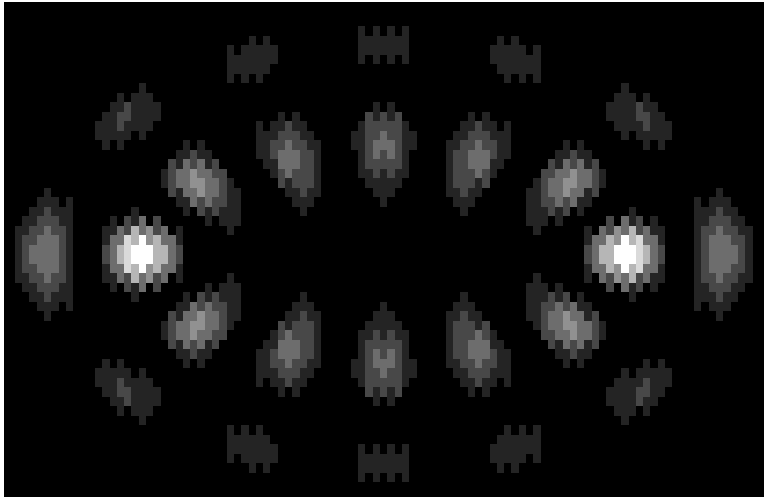
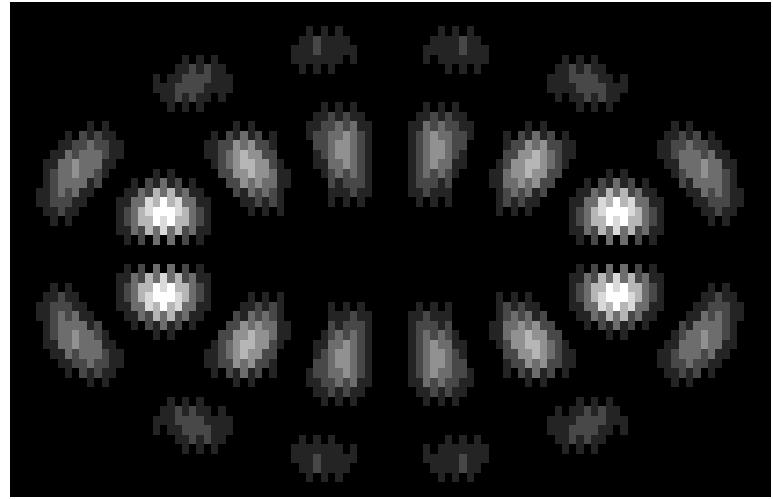
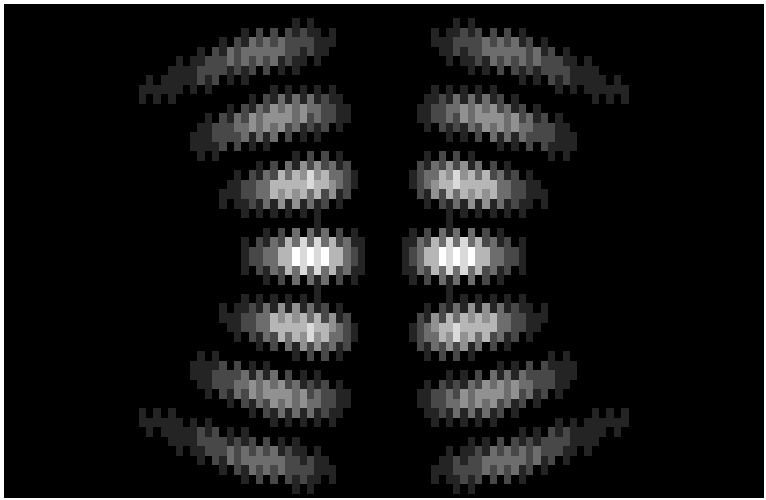
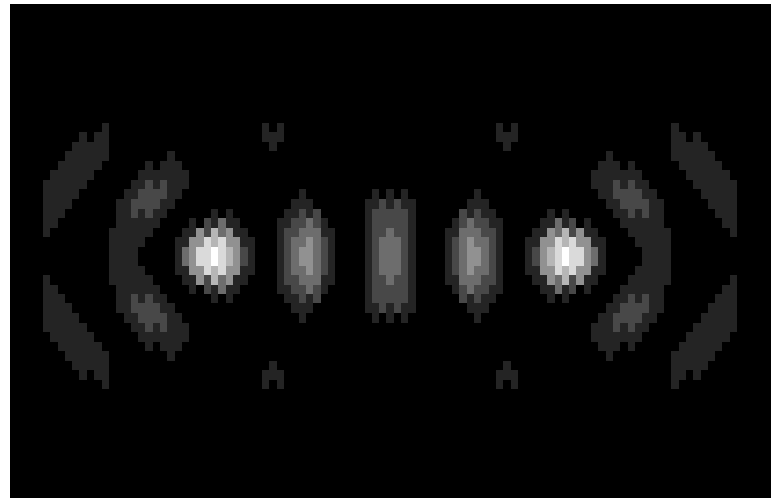
\end{document}